\documentclass{aastex}
\usepackage{spr-astr-addons}
\usepackage{url}

\RequirePackage{color}

\newcommand{\be}{\begin{equation}} \newcommand{\ee}{\end{equation}}
\newcommand{\bea}{\begin{eqnarray}} \newcommand{\eea}{\end{eqnarray}}
\newcommand{\bse}{\begin{subequations}}\newcommand{\ese}{\end{subequations}}
\newcommand{\n}{\nonumber}

\begin{document}

\title{Electromagnetic and anisotropic extension of a plethora of well-known solutions describing relativistic compact objects}
%% Running heads

\author{K. Komathiraj}
\affil{Department of Mathematical Sciences, Faculty of Applied Sciences, South Eastern University of Sri Lanka, Sammanthurai 3000, Sri Lanka.\\
Email: komathiraj@seu.ac.lk}

\and

\author{Ranjan Sharma}
\affil{Department of Physics, Cooch Behar Panchanan Barma University, Cooch Behar 736101, West Bengal, India.\\
Email: rsharma@associates.iucaa.in}

\shorttitle{Electromagnetic and anisotropic generalization}
\shortauthors{Komathiraj and Sharma}

\begin{abstract}
We demonstrate a technique to generate new class of exact solutions to the Einstein-Maxwell system describing a static spherically symmetric relativistic star with anisotropic matter distribution. An interesting feature of the new class of solutions is that one can easily switch off the electric and/or anisotropic effects in this formulation. Consequently, we show that a plethora of well known stellar solutions can be identified as sub-class of our class of solutions.  We demonstrate that it is possible to express our class of solutions in a simple closed form so as to examine its physical viability for the studies of relativistic compact stars.
\end{abstract}

\section{Introduction}
\label{sec:1}

Exact solutions to Einstein-Maxwell system play a major role in the studies of relativistic compact objects. While the Reissner-Nordstr\"om solution uniquely describes the exterior gravitational field of a static spherically isolated object in the presence of an electromagnetic field, a large class of interior solutions are available in the literature which are regular, well behaved and physically meaningful. In the uncharged case, a large class of such exact solutions and their physical viability have been examined by  \citet{Delgaty}. In the charged case, \citet{Ivanov} has compiled different class of exact solutions. 

In the recent past many new exact solutions have been developed some of which are, in fact, generalizations of many of the well-known solutions. Most of the extensions have generally been done either by incorporating an electromagnetic field or anisotropy or both into the system. The generalized models allow us to study the impacts of charge and/or anisotropy on the gross physical behaviour of a compact star. A prime motivating factor for such a generalization in most of our previous works was to fine-tune the stellar observables like mass and radius. 

Local anisotropy, as indicated by many investigators in the past, plays a significant role in the studies of relativistic stellar objects \citep{Rud,Bow,Herrera97}. In a recent article, it has been argued that pressure anisotropy cannot be ignored in the studies of relativistic compact stars as it is usually expected to develop by the physical processes inside such ultra-compact stars \citep{Herrera2020}. Incorporation of an electromagnetic field in the studies of astrophysical objects is also well-motivated and many pioneering works have been done in this field in the past which includes the pioneering works of \citet{Majumdar}, \citet{Papa}, \citet{Cooper} and \citet{Beken}, amongst others. Consequently, different stellar models have been developed by relaxing the pressure isotropy condition as well as by incorporating a net charge into the system. \cite{Sharma2001} have generalized the widely used \citet{Vaidya} stellar model by assuming a particular form of the electric field. The investigation shows a wide range of causal behaviour in the presence of the electric field. The Vaidya and Tikekar model was generalized by \citet{Karmakar} to analyze the impact of anisotropy on the maximum mass of a compact star. An anisotropic generalization of the Vaidya and Tikekar stellar model has been made by \citet{Thiru2019} recently. Earlier, \citet{Thiru2018} developed an algorithm to generalize a plethora of well-known exact solutions to Einstein field equations corresponding to a static spherically symmetric star by relaxing the pressure isotropy condition. \citet{Koma2018} have developed a formalism to generate a new class of interior solutions corresponding to the exterior R-N metric which contained many previously found solutions. \citet{Kom} also made an electromagnetic generalization of the \citet{Durgapal} stellar solution. By relaxing the pressure isotropy condition, \citet{Sharma2017} generalized the \citet{Finch} stellar model. For a specific charge distribution, \citet{Ratanpal2017} also made a generalization of the Finch and Skea stellar model to analyze the impact of the charge on the mass-radius relationship of a compact star, in particular. The relativistic stellar model of \citet{Mak2004} was extended by \citet{Kom2007} to include charge into the system. \citet{Maharaj2014} made a further generalization of the model by considering the system to be anisotropic as well. It is noteworthy that many of the static spherically symmetric anisotropic and/or charged stellar solutions available in the literature do not possess isotropic and/or charge neutral limits. 

In this paper, we intend to generate new class of exact solutions corresponding to a static spherically anisotropic star possessing a net charge. The idea is that once the anisotropy and/or charge are/is switched off we should be able to regain some of the well-behaved, physically interesting stellar solutions found earlier. Such a generalization would allow us to investigate the impacts of anisotropy and charge on the physical features of a compact object in a neat manner. Moreover, physical acceptability of the generalized solutions can be ensured by suitable choice of the anisotropic and/or charge parameters as their isotropic and uncharged counterparts have already been found to be regular, well behaved and physically meaningful.  

The paper is organized as follows: In Section \ref{sec:2}, we lay down the Einstein-Maxwell equations for an anisotropic fluid distribution. In Section \ref{sec:3}, we propose a technique to generate solutions to the system of equations. We show how a large class of well known solutions can be regained from our general class of solutions. In Section \ref{sec:4}, we express our solution in a closed form to analyze its features and physical viability. We conclude by discussing the key results of our investigation in Section \ref{sec:5}.

\section{Spacetime metric and field equations}
\label{sec:2}

We write the interior of a static spherically symmetric star by the line element
\be \label{eq:1}
ds^{2} = -e^{2\nu(r)}dt^{2}+e^{2\lambda(r)}dr^{2}+r^{2}(d\theta^{2}+\sin^{2}\theta
d\phi^{2}),\ee
in coordinates $(x^{a})=(t,r,\theta,\phi)$, where $\nu(r)$  and $\lambda(r)$ are two unknown  functions. For an anisotropic fluid in the presence of an electromagnetic field, we assume the energy momentum tensor in the form
\be\label{eq:2} T^i_{j}=\textrm{diag}(-\rho-\frac{1}{2}E^{2}, p_{r}-\frac{1}{2}E^{2}, p_{t}+\frac{1}{2}E^{2}, p_{t}+\frac{1}{2}E^{2}),\ee
where $\rho$ is the energy density, $p_{r}$ is the radial pressure and $p_{t}$ is the tangential  pressure; measured relative to the comoving fluid $4$-velocity $u^{i}=e^{-\nu}\delta^{i}_{0}$. 

For the line element (\ref{eq:1}) and matter distribution (\ref{eq:2}), the Einstein field equations are obtained as
\bea\label{eq:3}
\label{eq:3a}\frac{1}{r^{2}}(1-e^{-2\lambda})+\frac{2\lambda^\prime}{r}e^{-2\lambda}&=&\rho+\frac{1}{2}E^{2},\\
\label{eq:3b}-\frac{1}{r^{2}}(1-e^{-2\lambda})+\frac{2\nu^\prime}{r}e^{-2\lambda}&=&p_{r}-\frac{1}{2}E^{2},\\
\label{eq:3c}e^{-2\lambda}\left(\nu^{\prime\prime}+{\nu^\prime}^2+\frac{\nu^\prime}{r}-\nu^\prime\lambda^\prime-\frac{\lambda^\prime}{r}\right)&=&p_{t}+\frac{1}{2}E^{2},\\
\label{eq:3d}\Delta&=&p_{t}-p_{r},\\
\label{eq:3e}\sigma&=&\frac{1}{r^{2}}e^{-\lambda}(r^{2}E)^\prime,
\eea  
where a prime ($\prime$) denotes derivative with respect to the radial coordinate $r$. In the above,  $E$ is the electric field, $\sigma$ is the charge density and $\Delta$ is the  measure of anisotropy or anisotropic factor.  We shall use units having $8\pi G = 1 = c$. 

The mass of the gravitating object within a stellar radius $r$ is defined as
\be \label{eq:4}m(r)=\frac{1}{2}\int_{0}^{r}\tilde{r}^{2}\rho(\tilde{r})d\tilde{r}.\ee

Solutions of the above equations determine the physical behaviour of the anisotropic fluid distribution. The system (\ref{eq:3a})-(\ref{eq:3e}) comprises five equations in eight unknowns namely, $\nu,~\lambda,~\rho,~p_{r},~p_{t},~E,~\Delta$ and $\sigma$. Therefore, it is necessary to choose any three of these variables involved in the integration process to solve the system.

\section {Generating new solutions}
\label{sec:3}

Solutions to the system can be obtained by making physically reasonable choices for any three of the independent variables. Accordingly, we begin by assuming a particular form for one of the gravitational potentials as
\be\label{eq:5} e^{2\lambda(r)}=\frac{1-ar^{2}/ R^{2}}{1-br^{2}/R^{2}},\ee
where $a ~\textrm{and }~ b$ are nonzero arbitrary constants and $R$ is the boundary of the star. A similar form of the metric potential was earlier used by \citet{Nas} for the modelling of a neutron star and also by \citet{Koma2018} for a superdense charged star. For particular choices of the parameters $a$ and $b$, it is possible to identify the metric ansatz with the following solutions: (i) charged stellar model of \citet{KoM1} for $b=1$; (ii) stellar model of developed by \citet{MaL} for $b=1$; (iii) superdense stellar model developed by \citet{Tik} for $a=-7, ~b=1$; (iv) Vaidya and Tikekar superdense stellar model \citep{Vaidya} for $a=-2, ~b=1$ and (v) Durgapal and Bannerji neutron star model \citep{DuB} for $a=1, ~b=1/2$. In other words, the general form (\ref{eq:5}) contains a large class of metric potentials which have been used to develop physically acceptable relativistic stellar models. 

Using (\ref{eq:3b}) and (\ref{eq:3c}), for the particular choice (\ref{eq:5}), we obtain
 \bea \label{eq:6} \left(1-\frac{ar^{2}}{R^{2}}\right)\left(1-\frac{br^{2}}{R^{2}}\right)\left(\nu^{\prime\prime}+{\nu^\prime}^2-\frac{\nu^\prime}{r}\right)\n\\
 - (b-a)\left(\frac{r}{R^{2}}\right)\left(\nu^{\prime}+\frac{1}{r}\right)+\frac{b-a}{R^{2}}\left(1-\frac{ar^{2}}{R^{2}}\right)\n\\
-\left(1-\frac{ar^{2}}{R^{2}}\right)^{2}E^{2}=\left(1-\frac{ar^{2}}{R^{2}}\right)^{2}\Delta,\eea
which is a highly non linear differential equation with more than one unknowns. To make the equation tractable, it is convenient at this stage to introduce the following transformation
\be\label{eq:7}
e^{\nu(r)}=\psi(x),~~~x^{2}=1-\frac{br^{2}}{R^{2}}.\ee
The above transformation helps us to simplify the integration procedure as demonstrated by \citet{MaL} and \citet{KoM1,KoM2}, amongst others.

Under the transformation (\ref{eq:7}), equation (\ref{eq:6}) becomes
\bea \label{eq:8} (b-a+ax^{2})\ddot{\psi}-ax\dot{\psi}+\n\\
\left[\frac{a(a-b)}{b}+\frac{R^{2}(b-a+ax^{2})^{2}}{b^{2}(x^{2}-1)}(E^{2}+\Delta)\right]\psi=0,\eea
in terms of the new dependent and independent variables $\psi$ and $x$ where a dot (.) denotes differentiation with respect to $x$.
Under the transformation, the system (\ref{eq:3}) takes the following equivalent form
\bea\label{eq:9}\label{eq:9a} \rho &=&\frac {b(b-a)}{R^{2}}\frac{(3b-a+ax^{2})}{(b-a+ax^{2})^{2}}-\frac{1}{2}E^{2},\\
\label{eq:9b}p_{r}&=&\frac{b}{R^{2}(b-a+ax^{2})}\left(-2bx\frac{\dot{\psi}}{\psi}+a-b\right)+\frac{1}{2}E^{2},\\
\label{eq:9c}p_{t}&=&p_{r}+\Delta,\\
\label{eq:9d}\sigma^{2}&=&\frac{b^{2}[2xE-(1-x^{2})\dot{E}]^{2}}{R^{2}(1-x^{2})(b-a+ax^{2})}.
\eea
The mass function (\ref{eq:4}) becomes
\be\label{eq:10} m(x)=-\frac{R^{3}}{2b^{\frac{3}{2}}}\int_{1}^{x} x\sqrt{1-x^{2}}\rho(x)dx,\ee
 in terms of the new variable $x$. 

Thus, we have essentially reduced the solution of the field equations to integrating equation (\ref{eq:8}). The differential equation (\ref{eq:8})  may be integrated once the electric field $E^{2}$ and the anisotropic factor $\Delta$ are known. Since we have the freedom to choose two more variables, we assume particular forms of the electric field $E^{2}$ and anisotropic parameter $\Delta$ at this stage.  It is noteworthy that though a variety of choices are possible, the choices must ensure that they are regular, well behaved and can generate physically plausible stellar models. Keeping this in mind, we choose
\bea
\label{eq:11}E^{2}(x)&=&\frac{\alpha ab(x^{2}-1)}{R^{2}(b-a+ax^{2})^{2}},\\
\label{eq:12}\Delta(x)&=&\frac{\beta ab(a-b)(x^{2}-1)}{R^{2}(b-a+ax^{2})^{2}},\eea
where $\alpha$ and $\beta$ are real constants. It should be pointed out that both the electric field and the anisotropic factor given in (\ref{eq:11}) and  (\ref{eq:12}) are regular at the centre of the star. Our plan is to make use of these assumptions together with the potential (\ref{eq:5}) to generate new class of solutions describing a stellar configuration with desirable physical features. Using (\ref{eq:11}) and  (\ref{eq:12}), we express (\ref{eq:8}) in the form
 \be \label{eq:13}(b-a+ax^{2})\ddot{\psi}-ax\dot{\psi}+\frac{a}{b}[(1+\beta)(a-b)+\alpha]\psi=0,\ee
which is the master equation of the system and has to be integrated to find an exact model for a charged sphere with anisotropic pressure. Note that an uncharged and isotropic stellar solution can be regained simply by switching off the charge parameter $\alpha=0$ and the anisotropic parameter $\beta=0$ in (\ref{eq:13}). We intend to find new class of solutions for $\alpha \neq 0$ and $\beta \neq 0$.

\subsection {New class of charged anisotropic stellar solutions}
In this section, we provide systematically a rich family of solutions to Einstein-Maxwell system describing an anisotropic charged superdense star in line with some of the previous treatments \citep{Sharma2017,ThM,MaT,MaK,KoM3}. We note that that the point $x=0$ is a regular singular point of the differential equation (\ref{eq:13}). Therefore, the solution of the differential equation (\ref{eq:13}) can be written in the form of an infinite series by the method of Frobenius:
\be \label{eq:14}\psi(x)=\sum_{i=0}^{\infty}c_{i}x^{i},\ee
where $c_{i}$ are the coefficients of the series. 

To complete the solution we need to find the coefficients $c_{i}$  explicitly. Substituting (\ref{eq:14}) in (\ref{eq:13}), we obtain the recurrence relation
\bea\label{eq:15}a\left[\frac{1}{b}[(1+\beta)(a-b)+\alpha]+(i-2)(i-4)\right]c_{i-2}\n\\
+(b-a)i(i-1)c_{i}=0,  i\geq 2\eea
which governs the structure of the solution. With the help of (\ref{eq:15}), we express the general form for the even coefficients and odd coefficients in terms of the leading coefficient $c_{0}$ and $c_{1}$ respectively as:
\bea \label{eq:16}c_{2i}&=&\left(\frac{a}{a-b}\right)^{i}\frac{1}{(2i)!}\prod_{k=1}^{i}\left[\frac{1}{b}[(a-b)(1+\beta)+\alpha]\right.\n\\
&&\left.+(2k-2)(2k-4)\right]c_{0}\\ 
\label{eq:17}c_{2i+1}&=&\left(\frac{a}{a-b}\right)^{i}\frac{1}{(2i+1)!}\prod_{k=1}^{i}\left[\frac{1}{b}[(a-b)(1+\beta)+\alpha]\right.\n\\
&&\left.+(2k-1)(2k-3)\right]c_{1}.\eea
It is possible to  verify the results (\ref{eq:16}) and (\ref{eq:17}) by using mathematical induction.

Using (\ref{eq:14}), (\ref{eq:16}) and (\ref{eq:17}), we can now generate the general  solutions to (\ref{eq:13}), for the choice of the electric field (\ref{eq:11}) and the anisotropic factor (\ref{eq:12}), as
\be \label{eq:18}\psi(x)=c_{0}\psi_{1}(x)+c_{1}\psi_{2}(x),\ee
where we have set
\bea\label{eq:19}&&1+\sum_{i=1}^{\infty}\left(\frac{a}{a-b}\right)^{i}\frac{1}{(2i)!}\prod_{k=1}^{i}\left[\frac{1}{b}[(a-b)(1+\beta)+\alpha]\right.\n\\
&&\left.+(2k-2)(2k-4)\right]x^{2i}=\psi_{1}(x)\\
\label{eq:20}&& x+\sum_{i=1}^{\infty}\left(\frac{a}{a-b}\right)^{i}\frac{1}{(2i+1)!}\prod_{k=1}^{i}\left[\frac{1}{b}[(a-b)(1+\beta)+\alpha]\right.\n\\
&&\left.+(2k-1)(2k-3)\right]x^{2i+1}=\psi_{2}(x).\eea
The general solution (\ref{eq:18})  can be expressed in terms of elementary functions which is a more desirable form for the physical description of a charged anisotropic relativistic star. This is possible, in general, because the series (\ref{eq:19})  and (\ref{eq:20}) terminate for restricted values of the parameters $a,~b,~\alpha$ and  $\beta$ so that elementary functions are possible. 

In our work, we develop two sets of general solutions in terms of elementary functions by imposing the specific restrictions on the quantity $\frac{1}{b}[(a-b)(1+\beta)+\alpha]$ for a terminating series. The elementary functions, obtained using this method, can be written as polynomials and polynomials with algebraic functions.

We express the first category of solutions to (\ref{eq:13})  as
\bea \label{eq:21}\psi(x)&=&A\sum_{i=0}^{n}\gamma^{i}\frac{(n+i-2)!}{(n-i)!(2i)!}x^{2i}+B(b-a+ax^{2})^{3/2}\times\n\\
&&\sum_{i=0}^{n-2}\gamma^{i}\frac{(n+i)!}{(n-i-2)!(2i+1)!}x^{2i+1},\eea
where \bea \gamma &=&\frac{4a[4n(1-n)+1+\beta]}{4an(n-1)+\alpha},\n\\
4n(1-n)&=&\frac{1}{b}[(a-b)(1+\beta)+\alpha],\n\\
x^{2}&=&1-\frac{a(1+\beta)+\alpha}{4n(1-n)+1+\beta}\frac{r^{2}}{R^{2}}.\n\eea 

The second category of solution is obtained as
\bea\label{eq:22} \psi(x)&=& A \sum_{i=0}^{n} \mu^{i}\frac{(n+i-1)!}{(n-i)!(2i+1)!}x^{2i+1}\n\\
&+&B (b-a+ax^{2})^{3/2}\sum_{i=0}^{n-1} \mu^{i}\frac{(n+i)!}{(n-i-1)!(2i)!}x^{2i},\eea
where \bea \mu &=&\frac{4a(2-4n^{2}+\beta)}{a(4n^{2}-1)+\alpha},\n\\
1-4n^{2}&=&\frac{1}{b}[(a-b)(1+\beta)+\alpha],\n\\
x^{2}&=&1-\frac{a(1+\beta)+\alpha}{2-4n^{2}+\beta}\frac{r^{2}}{R^{2}}.\n\eea
In the above, $A$ and $B$ are arbitrary constants. 

Thus, we generate two new class of solutions (\ref{eq:21}) and (\ref{eq:22}) in terms of elementary functions from the infinite series solution (\ref{eq:18}). It should be stressed that the new class of solutions holds good for isotropic as well as anisotropic; charged as well as uncharged cases.  In the following, we demonstrate how our class of solutions can be used to regain a wide variety of previously developed well known stellar solutions:

\subsection {Sub-class of solutions}
It is not difficult to show that a large class of previously developed stellar models are actually sub-classes of our general class of solutions. The known solutions can either be explicitly regained from the general series solution (\ref{eq:18}) or from the elementary functions (\ref{eq:21}) and (\ref{eq:22}). This is illustrated by generating the following stellar models:

\subsubsection {The anisotropic and uncharged stellar model of \citet{Nas} }
{\bf Class: I}\\
We set $a=-\tilde{b},~b=(\tilde{b}-k)/2,~\alpha=0,$ and $\beta=2k/(k-3\tilde{b})~ (n=1)$ in (\ref{eq:22}) so that $\mu=8\tilde{b}/(k-3\tilde{b})$. Equation (\ref{eq:22}) then takes the form
\bea \label{eq:26}\psi(x)&=&A_{1}\sqrt{\tilde{b}(2+(k-\tilde{b})\tilde{x})}(5\tilde{b}-3k+2\tilde{b}(\tilde{b}-k)\tilde{x})\n\\
&+&B_{1}(1+\tilde{b}\tilde{x})^{3/2},\eea
where, $ x^{2}=1-\frac{a(1+\beta)}{\beta-2}\frac{r^{2}}{R^{2}} = 1+\frac{(k-\tilde{b})}{2}\tilde{x},~\tilde{x}=Cr^{2}(=\frac{r^{2}}{R^{2}}),$ and $A_{1}, B_{1}$ are constants.
The  solution (\ref{eq:26}) was the first of its class solutions found by \citet{Nas}. The exact solution (\ref{eq:26}) has been comprehensively studied \citep{Nas} and it has been shown the solution corresponds to an uncharged and anisotropic fluid sphere satisfying all the necessary conditions of a physically acceptable stellar model.

{\bf Class: II}\\
We set $a=-\tilde{b},~b=(\tilde{b}-k)/7,~\alpha=0,$~ and ~$\beta=7k/(k-8\tilde{b})~(n=2)$~ in~ (\ref{eq:21}),~ so that~ $\gamma=28\tilde{b}/(k-3\tilde{b})$. Subsequently, equation (\ref{eq:21}) becomes
\bea \label{eq:27}\psi(x)&=&A_{2}[3(k-8\tilde{b})^{2}+12\tilde{b}(k-8\tilde{b})(7+(k-\tilde{b})\tilde{x})\n\\
&+&8\tilde{b}^{2}(7+(k-\tilde{b})\tilde{x})^{2}]\n\\
&+&B_{2}(1+\tilde{b}\tilde{x})^{3/2}\sqrt{7\tilde{b}-\tilde{b}(\tilde{b}-k)\tilde{x}},\eea
where, $ x^{2}=1-\frac{a(1+\beta)}{\beta-7}\frac{r^{2}}{R^{2}}=1+\frac{(k-\tilde{b})}{7}\tilde{x},\\
~\tilde{x}=Cr^{2}(=\frac{r^{2}}{R^{2}}),$  and $A_{2}, B_{2}$ are constants. The solution (\ref{eq:27}) was the second class of solutions obtained by \citet{Nas}. Note that our solution (\ref{eq:27}) corrects a minor misprint in the result obtained by \citep{Nas}.

\subsubsection {Charged stellar model of \citet{KoM1} }

If we set  $\beta=0$, equation (\ref{eq:21}) yields 
\bea \label{eq:28}\psi(x)&=&A\sum_{i=0}^{n}(-\gamma)^{i}\frac{(n+i-2)!}{(n-i)!(2i)!}x^{2i}\n\\
&+&B(b-a+ax^{2})^{3/2}\n\\
&\times&\sum_{i=0}^{n-2}(-\gamma)^{i}\frac{(n+i)!}{(n-i-2)!(2i+1)!}x^{2i+1},\eea
where $\gamma =4-\frac{4b}{4bn(n-1)+\alpha},~a+\alpha=b[2-(2n-1^{2})],\\~x^{2}=1-\frac{br^{2}}{R^{2}}.$

If we set  $\beta=0$, equation (\ref{eq:22}) yields

\bea\label{eq:29} \psi(x)&=& A \sum_{i=0}^{n} (-\mu)^{i}\frac{(n+i-1)!}{(n-i)!(2i+1)!}x^{2i+1}\n\\
&+&B (b-a+ax^{2})^{3/2}\n\\
&\times&\sum_{i=0}^{n-1} (-\mu)^{i}\frac{(n+i)!}{(n-i-1)!(2i)!}x^{2i},\eea
where $\mu =4-\frac{4b}{b(4n^{2}-1)+\alpha},~a+\alpha =2b(1-2n^{2}),\\~x^{2}=1-\frac{br^{2}}{R^{2}}$. 

Solutions (\ref{eq:28}) and (\ref{eq:29})  correspond to the isotropic charged stellar model of \citet{KoM2}. These solutions reduce to \citet{KoM1} model if we set $b=1.$

\subsubsection {Superdense stellar model of \citet{MaL} }

If we set  $b=1,~\alpha=0$ and $\beta=0$, then equation (\ref{eq:21}) yields
\bea \label{eq:30}\psi(x)&=&A\sum_{i=0}^{n}(-\gamma)^{i}\frac{(n+i-2)!}{(n-i)!(2i)!}x^{2i}+B(1-a+ax^{2})^{3/2}\n\\
&\times&\sum_{i=0}^{n-2}(-\gamma)^{i}\frac{(n+i)!}{(n-i-2)!(2i+1)!}x^{2i+1},\eea
where $\gamma =4-\frac{4}{4n(n-1)},~a=[2-(2n-1^{2})],~x^{2}=1-\frac{r^{2}}{R^{2}}.$

If we set  $b=1,~\alpha=0$ and $\beta=0$, then (\ref{eq:22}) yields

\bea \label{eq:31}\psi(x)= A \sum_{i=0}^{n} (-\mu)^{i}\frac{(n+i-1)!}{(n-i)!(2i+1)!}x^{2i+1}\n\\
+B (1-a+ax^{2})^{3/2}\sum_{i=0}^{n-1} (-\mu)^{i}\frac{(n+i)!}{(n-i-1)!(2i)!}x^{2i},\eea
where  $\mu =4-\frac{4}{(4n^{2}-1)},~a =2(1-2n^{2}),~x^{2}=1-\frac{r^{2}}{R^{2}}.$ 
 
These two categories of solutions (\ref{eq:30})  and (\ref{eq:31}) correspond to \citet{MaL} model describing a relativistic compact star. The Maharaj and Leach solution has a simple form in terms of elementary functions and provides a physically reasonable model for neutron stars.

\subsubsection {\citet{Tik} superdense stellar model}

If we set $a=-7,~b=1,~\alpha=0,$ and $\beta=0~ (n=2)$ in (\ref{eq:21}), then we get $\gamma=-\frac{7}{2}$ and subsequently equation (\ref{eq:21}) yields

\be \label{eq:32} \psi(x)=A_{3}\left(1-\frac{7}{2}x^{2}+\frac{49}{24}x^{4}\right)+B_{3}x\left(1-\frac{7}{8}x^{2}\right)^{3/2},\ee
where, $x^{2}=1-\frac{r^{2}}{R^{2}}$, $A_{3}$ and $B_{3}$ are constants. The solution (\ref{eq:32}) was found by \citet{Tik} for the description of compact stars like neutron stars. 

\subsubsection {\citet{Vaidya} compact stellar model}

If we set $a=-2,~b=1,~\alpha=0,$ and $\beta=0 ~(n=1)$ in (\ref{eq:22}), then we have $\mu=-\frac{8}{3}$ and equation (\ref{eq:22}) becomes

\be \label{eq:33}\psi(x)=A_{4}x\left(1-\frac{4}{9}x^{2}\right)+B_{4}\left(1-\frac{2}{3}x^{2}\right)^{3/2},\ee
where, $x^{2}=1-\frac{r^{2}}{R^{2}}$, $A_{4}$ and $B_{4}$ are constants. The exact solution (\ref{eq:33}), developed by \citet{Vaidya}, has been widely used for the studies of relativistic compact stars.

\subsubsection{\citet{DuB} relativistic stellar model}

If we set $a=-1,~b=\frac{1}{2},~\alpha=0,$ and $\beta=0 ~(n=1)$ in (\ref{eq:22}), then we have $\mu=-\frac{8}{3}$ and equation (\ref{eq:22}) yields

\be \label{eq:34}\psi(x)=A_{5}(2-\tilde{x})^{1/2}(5+2\tilde{x})+B_{5}(1+\tilde{x})^{3/2},\ee
where, $ x^{2}=1-\frac{1}{2}\frac{r^{2}}{R^{2}}=1-\frac{1}{2}\tilde{x},~\tilde{x}=Cr^{2}(=\frac{r^{2}}{R^{2}}),$\\ and $A_{5}$ and $B_{5}$ are constants which is the \citet{DuB} stellar model. The  model has been shown to satisfy all the physical requirements of a realistic star and has got widespread attention for the modeling of relativistic stellar configurations.

\section {New closed-form solutions and its physical features}
\label{sec:4}
In the previous section, it has been shown how a large class of previously reported solutions can be regained from our general class of solutions. It is interesting to note that the solutions can also be obtained in simple analytic forms which facilitates its physical analysis. This is demonstrated as follows.
  
We set $$b=\frac{a(1+\beta)+\alpha}{\beta-2}~(n=1)$$ in (\ref{eq:22}) so that we have $$\mu=\frac{4a(\beta-2)}{3a+\alpha}.$$
Equation (\ref{eq:22}) then yields
\bea \label{eq:23}\psi(x) &=& Ax\left(1+\frac{2a(\beta-2)}{3(3a+\alpha)}x^{2}\right)\n\\
&+&B\left(\frac{3a+\alpha}{\beta-2} +ax^{2}\right)^{3/2},\eea
where, $x^{2}=1-\frac{a(1+\beta)+\alpha}{\beta-2}\frac{r^{2}}{R^{2}}.$
Using (\ref{eq:5}), (\ref{eq:7}), (\ref{eq:9a})-(\ref{eq:9d}), (\ref{eq:11}) and (\ref{eq:12}), we obtain 
\bea\label{eq:24}\label{eq:24a}e^{2\lambda}&=&\frac{\alpha+3a+a(\beta-2)x^{2}}{(\alpha+a+a\beta)x^{2}},\\
\label{eq:24b}e^{2\nu}&=&\psi^{2},\\
\label{eq:24c} \rho &=&\frac{(3a+\alpha)(a+\alpha+a\beta)[3\alpha+a(5+2\beta+(\beta-2)x^{2})]}{R^{2}(\beta-2)[3a+\alpha+a(\beta-2)x^{2}]^{2}}\n\\
&-&\frac{\alpha a(\beta-2)(a+\alpha+a\beta)(x^{2}-1)}{2R^{2}[3a+\alpha+a(\beta-2)x^{2}]^{2}},\\
\label{eq:24d}p_{r}&=&-\frac{(a+\alpha+a\beta)}{R^{2}(\beta-2)[3a+\alpha+a(\beta-2)x^{2}]}\n\\
&\times&\left[3a+\alpha+2(a+\alpha+a\beta)x\frac{\dot{\psi}}{\psi}\right]\n\\
&+&\frac{\alpha a(\beta-2)(a+\alpha+a\beta)(x^{2}-1)}{2R^{2}[3a+\alpha+a(\beta-2)x^{2}]^{2}},\\
\label{eq:24e}p_{t}&=&p_{r}+\frac{\beta a(3a+\alpha)(a+\alpha+a\beta)(1-x^{2})}{R^{2}[3a+\alpha+a(\beta-2)x^{2}]^{2}},\\
\label{eq:24f}E^{2}&=&\frac{\alpha a(\beta-2)(a+\alpha+a\beta)(x^{2}-1)}{R^{2}[3a+\alpha+a(\beta-2)x^{2}]^{2}},\\
\label{eq:24g}\Delta&=&\frac{\beta a(3a+\alpha)(a+\alpha+a\beta)(1-x^{2})}{R^{2}[3a+\alpha+a(\beta-2)x^{2}]^{2}},\\
\label{eq:24h}\sigma^{2}&=&-\alpha ax^{2}(a+\alpha+a\beta)^{3}\n\\
&\times&\frac{[3\alpha+a(5+2\beta+(\beta-2)x^{2})]^{2}}{R^{4}[3a+\alpha+a(\beta-2)x^{2}]^{5}},
\eea
where $\psi$ is given in (\ref{eq:23}). 

The mass function(\ref{eq:10}) takes the form
\bea \label{eq:25} m(x)=-\frac{3\alpha R}{8(a)^{\frac{3}{2}}}\tanh^{-1}\left[\sqrt{\frac{a(\beta-2)(1-x^{2})}{(a+\alpha+a\beta)}}\right]\n\\
-\frac{R\sqrt{(\beta-2)(1-x^{2})}}{8a\sqrt{a+\alpha+a\beta}[3a+\alpha+a(\beta-2)x^{2}]}\n\\
\times[12a^{2}(x^{2}-1)-3\alpha^{2}-a\alpha(11+\beta+2(\beta-4)x^{2})].\eea

The simple closed-form nature of the above solution facilitates its physical analysis as discussed below.

We note from (\ref{eq:24a}) that \\
$e^{2\lambda}(r=0)=1,~~(e^{2\lambda})'(r=0)=0$\\
and from (\ref{eq:24b}) we have \\
$e^{2\nu}(r=0)=\left[A\left(1+\frac{2a(\beta-2)}{3(3a+\alpha)}\right)+B\left(\frac{3a+\alpha}{\beta-2} +a\right)^{3/2}\right]^{2},\\
~~(e^{2\nu})'(r=0)=0$.\\
Obviously, the gravitational potentials are regular at the origin. 

From (\ref{eq:24c}), we obtain the central density as\\
$\rho_{0}=\rho(r=0)=\frac{3(3a+\alpha)}{R^{2}(\beta-2)},$\\
 which implies that we must have 
\be \label{eq:35}\frac{(3a+\alpha)}{(\beta-2)} > 0.\ee
Using (\ref{eq:24d}) and (\ref{eq:24e}), we obtain the radial and tangential pressures at $r=0$ as 
\bea p_{r}(r=0) &=& p_{t}(r=0) \n\\
&=& -\frac{3a+\alpha+2(a+\alpha+a\beta)\left[\frac{\dot{\psi}}{\psi}\right]_{r=0}}{R^{2}(\beta-2)},\n\eea
where $\psi$ is given in (\ref{eq:23}). That density and pressure should be positive puts the following bound on the model parameters
\bea \label{eq:36} 0 >\frac{3a+\alpha}{2-\beta} > \frac{2(a+\alpha+a\beta)}
{(\beta-2)\left(A\frac{5a+3\alpha+2a\beta}{3(3a+\alpha)}+B\left(\frac{a+\alpha+a\beta}{\beta-2}\right)^{\frac{3}{2}}\right)}\n\\
\times\left(A\frac{\alpha-a+2a\beta}{3a+\alpha}+3Ba\sqrt{\frac{a+\alpha+a\beta}{\beta-2}}\right).\eea

At the boundary of the star $(r=R)$, the radial pressure must vanish, i.e. $$p_{r}(r=R)=p_{r}(x=\sqrt{1-\frac{a(1+\beta)+\alpha}{\beta-2}})=0.$$
which yields
\bea\label{eq:37} \frac{B}{A}=\sqrt{\frac{2+a+\alpha-\beta(1-a)}{2-\beta}} \left(\frac{\beta-2}{a+\alpha+a\beta}\right)^{\frac{3}{2}}\times\frac{s_1}{s_2},
\eea
where $s_1 = 12(1-a)(a+\alpha+p\beta)[-\alpha+a(1+2\alpha-2\beta+2a(1+\beta))]-[2\alpha+a(6-6a+\alpha(\beta-4))][3\alpha+a(5-2\alpha+2\beta-2a(1+\beta))]$ and  $s_2=3(a-1)(3a+\alpha)[-2\alpha+2a(9+9a+8\alpha)+a\beta(-12+12a-\alpha)].$

The exterior solution to the Einstein-Maxwell system for $r > R$ is given by the Reissner-Nordstr\"om line element
\bea\label{eq:38} ds^{2} &=& -\left(1-\frac{2M}{r}+\frac{Q^{2}}{r^{2}}\right)dt^{2}+\left(1-\frac{2M}{r}+\frac{Q^{2}}{r^{2}}\right)^{-1}dr^{2}\n\\
&+&r^{2}(d\theta^{2}+\sin^{2}\theta d\phi^{2},\eea
where, $M$ and $Q$ are the total mass and charge, respectively. Matching of the line element (\ref{eq:1}) and (\ref{eq:38}), across the boundary $r=R$, we have
\bea \left(1-\frac{2M}{R}+\frac{Q^{2}}{R^{2}}\right)^{-1}=\frac{(a-1)(\beta-2)}{3a+\alpha+(a-1)(\beta-2)},\\
\left(1-\frac{2M}{R}+\frac{Q^{2}}{R^{2}}\right)^{\frac{1}{2}}= A\sqrt{\frac{2+a+\alpha+(a-1)\beta}{2-\beta}}\n\\
\times\frac{[3\alpha+a(5-2\alpha+2\beta-2a(1+\beta))]}{3(3a+\alpha)}\n\\
+B\left(\frac{(1-a)(a+\alpha+a\beta)}{\beta-2}\right)^{\frac{3}{2}}.\eea
These matching conditions and (\ref{eq:37}) help us to determine the constants $A$ and $B$ explicitly in terms of the model parameters $a,~\alpha$ and $\beta$ as follows:
\bea \label{eq:41}A &=& \frac{f(a,\alpha,\beta)}{g(a,\alpha,\beta)},\\
\label{eq:42}B &=& \frac{h(a,\alpha,\beta)}{i(a,\alpha,\beta)},\eea
where,
\bea f(a,\alpha,\beta)&=&-\sqrt{1-\frac{3a+\alpha}{(1-a)(\beta-2)}}\times[2\alpha-2a(9+8\alpha)\n\\
&+&a\beta(12+\alpha)-6a^{2}(3+2\beta)],\n\\
g(a,\alpha,\beta)&=&4(3a+\alpha)\sqrt{\frac{2+a+\alpha-(1-a)\beta}{2-\beta}},\n\\
h(a,\alpha,\beta)&=&\sqrt{1-\frac{3a+\alpha}{(1-a)(\beta-2)}}\times\{[-2\alpha\n\\
&+&a(-6+6a-\alpha(\beta-4))]\n\\
&\times&[-3\alpha+a(-5+2\alpha-2\beta+2a(1+\beta))]\n\\
&+&12(a-1)(a+\alpha+p\beta)\n\\
&\times&[-\alpha+a(1+2\alpha-2\beta+2a(1+\beta))]\},\n\\
i(a,\alpha,\beta)&=&12(3a+\alpha)^{2}\left[\frac{(1-a)(a+\alpha+a\beta)}{\beta-2}\right]^{\frac{3}{2}}.\n\eea

\begin{figure}[tb]
\includegraphics[width=\columnwidth]{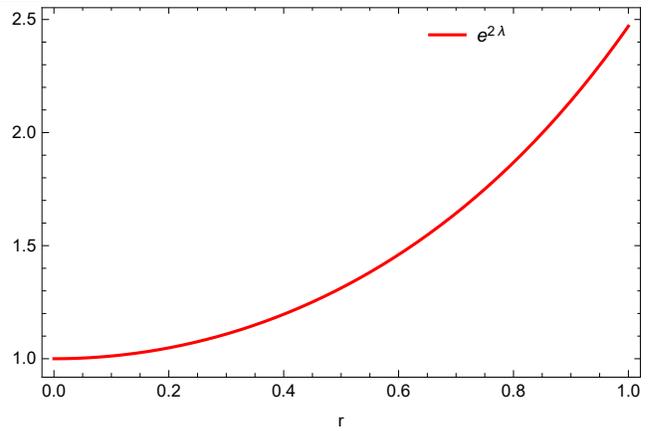}
\caption{Behavior of metric potential $e^{\lambda}$ within the stellar interior.}
\label{fig:1}  
\end{figure}

\begin{figure}[tb]
\includegraphics[width=\columnwidth]{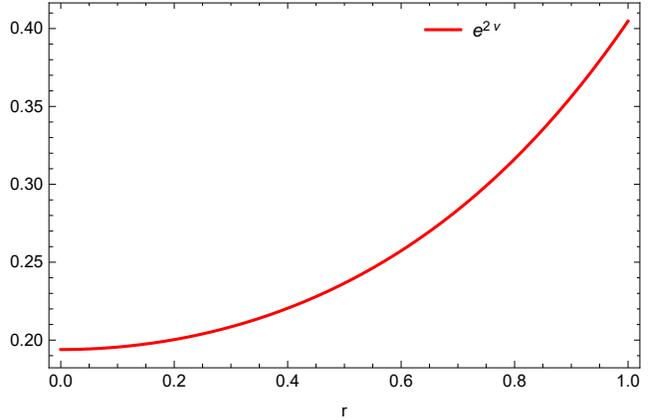}
\caption{Behavior of metric potential $e^{\nu}$ within the stellar interior.}
\label{fig:2}      
\end{figure}

\begin{figure}[tb]
\includegraphics[width=\columnwidth]{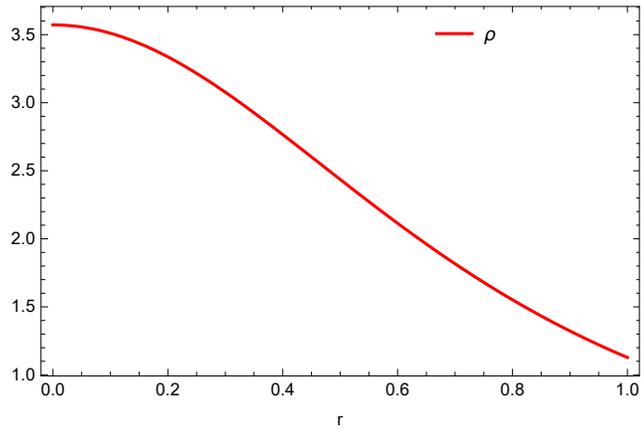}
\caption{Radial variation of energy density $\rho$.}
\label{fig:3}  
\end{figure}

\begin{figure}[tb]
\includegraphics[width=\columnwidth]{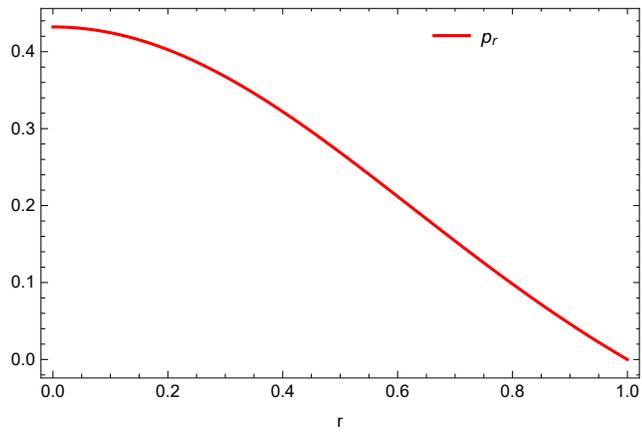}
\caption{Radial variation of radial pressure $p_r$.}
\label{fig:4}  
\end{figure}

\begin{figure}[tb]
\includegraphics[width=\columnwidth]{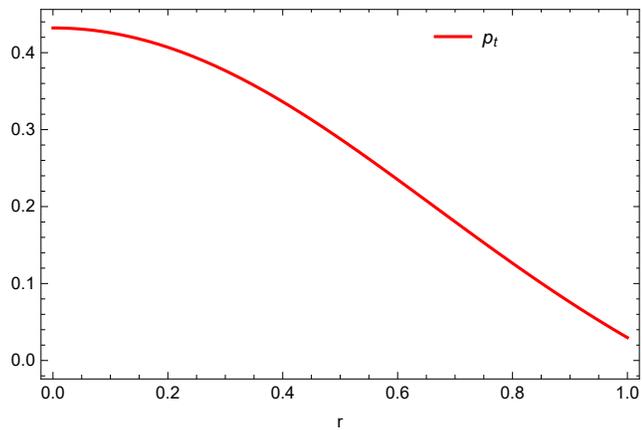}
\caption{Radial variation of tangential pressure $p_t$.}
\label{fig:5}  
\end{figure}

\begin{figure}[tb]
\includegraphics[width=\columnwidth]{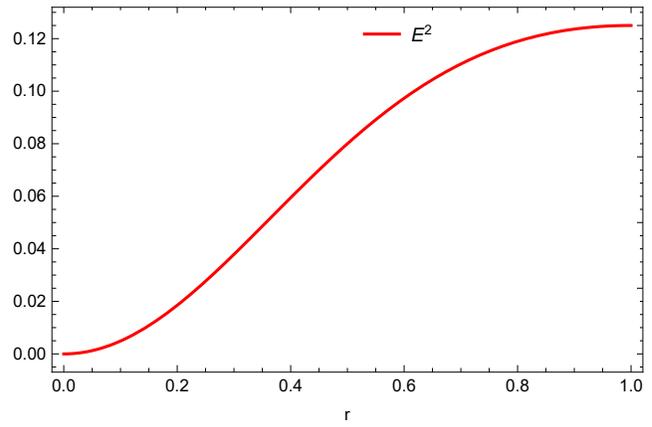}
\caption{Radial variation of electric field $E^2$.}
\label{fig:6}  
\end{figure}

\begin{figure}[tb]
\includegraphics[width=\columnwidth]{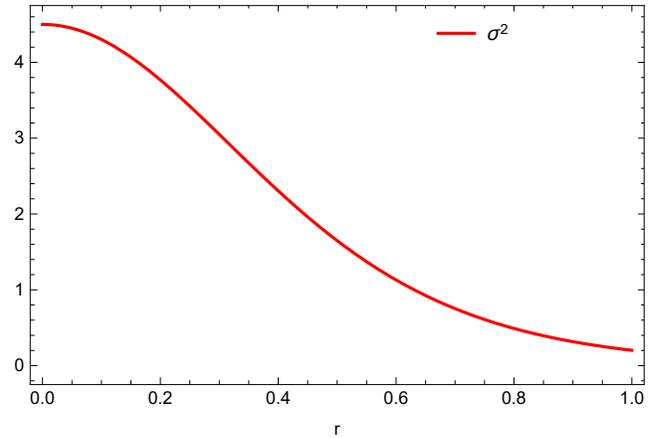}
\caption{Radial dependance of charge density $\sigma$.}
\label{fig:7}  
\end{figure}

\begin{figure}[tb]
\includegraphics[width=\columnwidth]{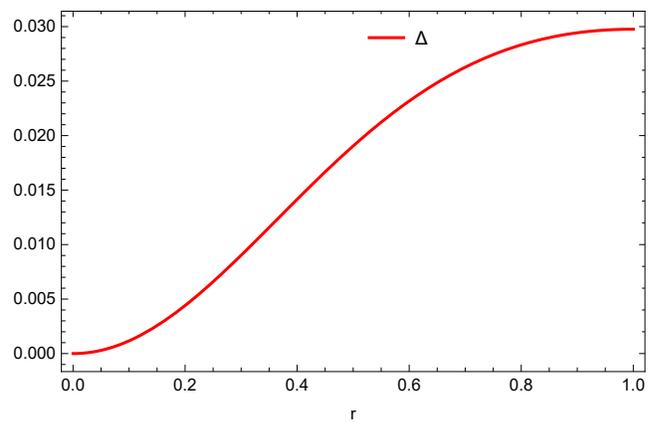}
\caption{Radial variation of anisotropy $\Delta$.}
\label{fig:8}  
\end{figure}

\begin{figure}[tb]
\includegraphics[width=\columnwidth]{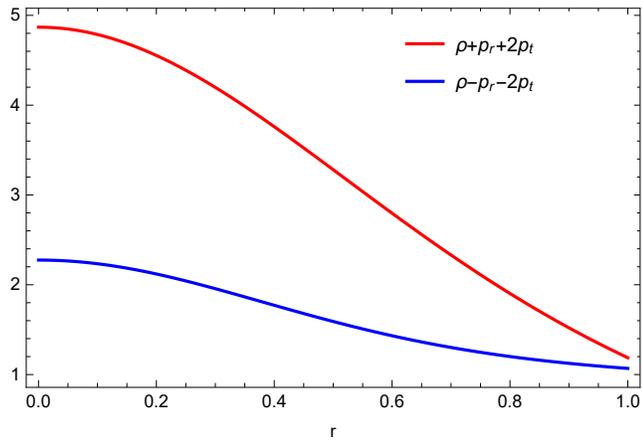}
\caption{Fulfillment of energy conditions.}
\label{fig:9}  
\end{figure}

\begin{figure}[tb]
\includegraphics[width=\columnwidth]{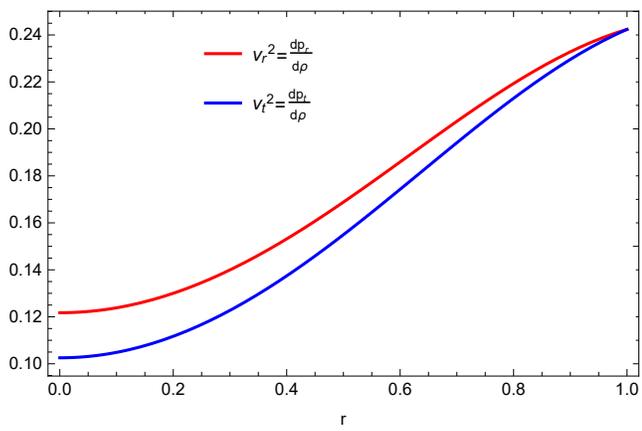}
\caption{Sound speeds within the stellar interior.}
\label{fig:10}  
\end{figure}

\begin{figure}[tb]
\includegraphics[width=\columnwidth]{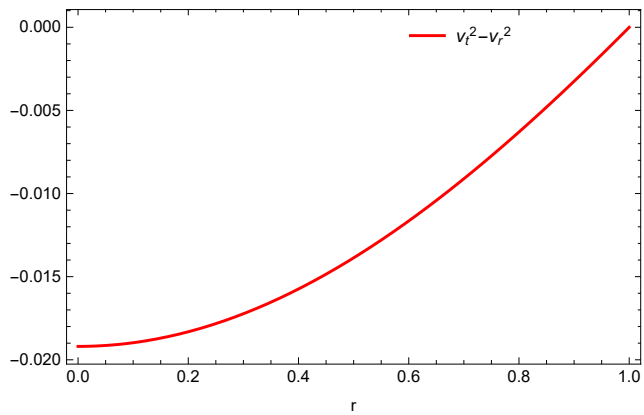}
\caption{Difference of sound speeds.}
\label{fig:11}  
\end{figure}

\begin{figure}[tb]
\includegraphics[width=\columnwidth]{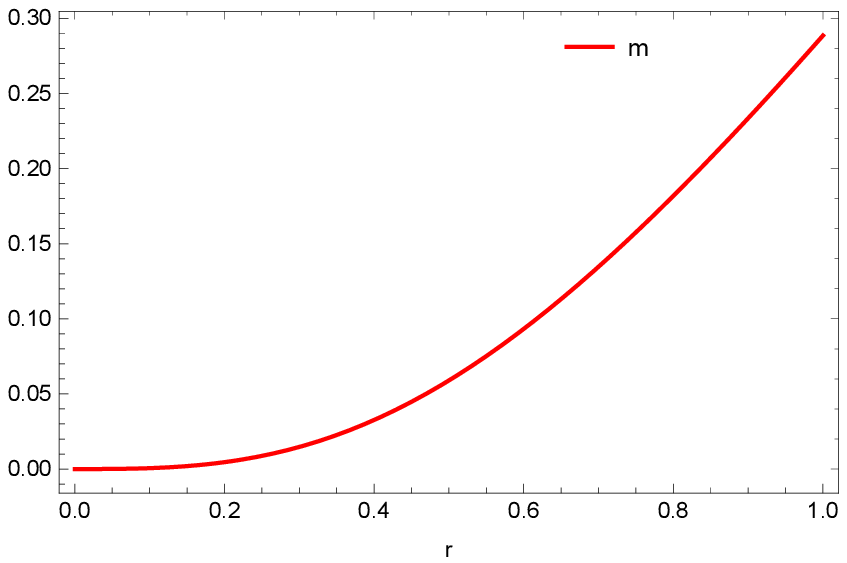}
\caption{Radial variation of the mass function $m(r)$.}
\label{fig:12}  
\end{figure}

To analyze behaviour of the physical variables, for a star of radius $R=1$ (with $8\pi G = c = 1$), we set $a = -1,~\alpha = 0.5$ and $\beta = -0.1$, which are consistent with the bounds (\ref{eq:35})-(\ref{eq:37}). Using these values in (\ref{eq:41}) and (\ref{eq:42}), we determine the constants as $A=0.809637$ and $B=1.01215.$ Making use of these values, we analyze the physical features of the model. Figures (\ref{fig:1}) and (\ref{fig:2}) show that the gravitational potentials $e^{2\lambda}$ and $e^{2\nu}$ are continuous, regular and well-behaved at the interior of the star. Figure (\ref{fig:3}) shows that the energy density $\rho$ is positive, finite and monotonically decreases radially outward from its maximum value at the centre. The behaviour of radial pressure $p_{r}$ and the tangential pressure $p_{t}$ are plotted in Figures in (\ref{fig:4}) and (\ref{fig:5}) respectively which show that both the pressures are positive and decreasing monotonically while the radial pressure vanishes at the boundary. In Figure (\ref{fig:6}), we show that the electric field intensity $E$ is continuous throughout the interior and increases from the centre to the boundary, which is physically reasonable. The radial variation of the charge density is shown in figure (\ref{fig:7}).  In Figure (\ref{fig:8}), radial variation of the anisotropic factor $\Delta$ is shown. We note that $\Delta$ is positive and monotonically increases from the centre until it attains a maximum value at the boundary of the star. This profile is similar to that obtained by \citet{MaM} and \citet{Kom}. Figure (\ref{fig:9}) illustrates that the energy conditions $\rho+p_{r}+2p_{t} > 0$ and $\rho-p_{r}-2p_{t} > 0$ are satisfied throughout the stellar configuration. In Figure (\ref{fig:10}), the sound speed in radial and transverse directions, i.e. $v_{r}^{2}=\frac{dp_{r}}{d\rho}$  and  $v_{t}^{2}=\frac{dp_{t}}{d\rho}$ are shown which confirms that the causality condition is not violated in this model, a desirable feature for the modelling of a stellar structure as pointed out by \citet{Delgaty}. For an anisotropic fluid sphere, a potentially stable configuration is ensured if we have $-1\leq v_{t}^{2}-v_{r}^{2}\leq 0$ \citep{Her,Abr}. This condition is also satisfied in our model as shown in Figure (\ref{fig:11}). Figure (\ref{fig:12}) shows the mass function profile within the stellar interior which is regular at the centre. Figure (\ref{fig:13}) shows the mass function profile within the stellar interior which is regular at the centre. Thus, we show that there exists particular sets of model parameters for which solution (\ref{eq:24}) satisfies all the requirements of a realistic star.

\section{Discussion}
\label{sec:5}

Through our investigation, we have provided a general class of charged anisotropic relativistic stellar solutions which is regular and well-behaved. The most interesting feature of the class of solutions is that many well known stellar solutions can be regained simply by switching off the parameters specifying the anisotropy and/or charge distribution in this formulation. 

It is to be stressed that for physical analysis, we have generated one particular closed form solution by suitably fixing the model parameters. It will be interesting to probe what other combinations of the model parameters can provide new solutions in simple analytic forms. This will be taken up elsewhere.

\acknowledgements
RS gratefully acknowledges support from the Inter-University Centre for Astronomy and Astrophysics (IUCAA), Pune, India, under its Visiting Research Associateship Programme.

\end{document}